\documentclass[12pt]{article}
\usepackage{times}
\usepackage{geometry}
\usepackage[utf8]{inputenc}
\usepackage{enumitem,amssymb}
\usepackage{ragged2e}

\usepackage{graphicx}
\usepackage{epsfig}
\usepackage[font=footnotesize,labelfont=bf]{caption}
\usepackage[rflt]{floatflt}
\usepackage{natbib}
\newlist{thematic}{itemize}{8}
\setlist[thematic]{label=$\square$}
\usepackage{pifont}
\newcommand{\cmark}{\ding{51}}%
\newcommand{\done}{\rlap{$\square$}{\raisebox{2pt}{\large\hspace{1pt}\cmark}}%
\hspace{-2.5pt}}

\begin{document}
\raggedright
\huge
Astro2020 Science White Paper \\[0.25ex]

Using X-Ray Polarimetry to Probe the Physics of Black Holes and Neutron Stars \\[1ex]
\normalsize

\noindent \textbf{Thematic Areas:} \hspace*{60pt} $\square$ Planetary Systems \hspace*{10pt} $\square$ Star and Planet Formation \hspace*{20pt}\linebreak
$\done$ Formation and Evolution of Compact Objects \hspace*{31pt} $\done$ Cosmology and Fundamental Physics \linebreak
  $\square$  Stars and Stellar Evolution \hspace*{1pt} $\square$ Resolved Stellar Populations and their Environments \hspace*{40pt} \linebreak
  $\square$    Galaxy Evolution   \hspace*{45pt} $\done$             Multi-Messenger Astronomy and Astrophysics \hspace*{65pt} \\[0.5ex]
  
\textbf{Principal Author:}

Name:	Henric S. Krawczynski
 \linebreak						
Institution:  Wash.\ U.\ St. Louis, MCSS \& QSC 
 \linebreak
Email: krawcz@wustl.edu
 \linebreak
Phone:  314 935 8553
 \\[0.25ex]
 
\textbf{Co-authors:} 
Giorgio Matt (U. Roma Tre),
Adam R. Ingram (Oxford), 
Roberto Taverna (U. Roma Tre, U. Padova),
Roberto Turolla (U. Padova, MSSL),
Fabian Kislat (UNH),
C. C. Teddy Cheung (NRL),
Andrei Bykov (Ioffe Inst., St. Petersburg),
Kuver Sinha (U. Oklahoma),
Haocheng Zhang (Purdue),
Jeremy Heyl (UBC),
Niccol\`o Bucciantini (INAF Arcetri),
Greg Madejski (SLAC, Stanford),
Tim Kallman (GSFC),
Keith M.\ Jahoda (GSFC),
Quin Abarr (Wash.\ U.\ St. Louis, MCSS), 
Matthew G.\ Baring (Rice),
Luca Baldini (U. Pisa, INFN),
Mitchell Begelman (CU Boulder),
Markus B\"ottcher (NWU),
Edward Cackett (Wayne State),
Ilaria Caiazzo (UBC),
Paolo Coppi (Yale),
Enrico Costa (INAF-IASF Roma),
Jason Dexter (CU Boulder),
Jean-Fran\c cois Fortin (U. Laval),
Charles Gammie (UI Urbana Champaign),
Jessica A.\ Gaskin (MSFC),
Dimitrios Giannios (Purdue),
Pranab Ghosh (Tata Inst.),
Alice K. Harding (GSFC),
Dieter H.\ Hartmann (Clemson),
Kiyoshi Hayashida (Osaka U.),
Svetlana G. Jorstad (Boston U., St.~Petersburg State U.),
Philip Kaaret (Univ. of Iowa),
Takao Kitaguchi (RIKEN),
Luca Latronico (Torino),
Thomas Maccarone (Texas Tech),
Alan Marscher (Boston U.),
Herman Marshall (MIT), 
Mark McConnell (UNH),
Jon M. Miller (UM Ann Arbor),
Stephen L.\ O'Dell (MSFC), 
Feryal \"Ozel (UA Tucson),
Takashi Okajima (GSFC),
Mark Pearce (KTH),
Jeremy Perkins (GSFC),
Brian D. Ramsey (MSFC),
Roger W. Romani (Stanford), 
Jeremy D. Schnittman (GSFC),
Carmelo Sgro (Pisa),
Paolo Soffitta (INAF-IASF Roma),
Hiromitsu Takahashi (Hiroshima U.),
Toru Tamagawa (RIKEN),
John Tomsick (UC Berkeley, SSL),
Martin C.\ Weisskopf (MSFC),
George Younes (GWU).\\[-0.1ex]
\hspace*{1cm}
\linebreak

\justify

\textbf{Abstract:}
This white paper highlights compact object and fundamental physics science opportunities afforded by high-throughput broadband (0.1-60 keV) X-ray polarization observations. 
X-ray polarimetry gives new observables 
with geometric information about stellar remnants which are many orders of magnitude 
too small for direct imaging.
The X-ray polarimetric data also reveal details
about the emission mechanisms and the structure of the magnetic fields in and around the most 
extreme objects in the Universe. 
Whereas the Imaging X-ray Polarimetry Explorer (IXPE) to be launched in 2021 will obtain first results for bright objects, a follow-up mission could be one order of magnitude more sensitive and would be able to use a broader bandpass to perform
physics type experiments for representative samples of sources. 
\thispagestyle{empty}
\clearpage
\justify
\setcounter{page}{1}
{\bf Introduction:} 
The recent developments of broadband X-ray focusing mirrors with excellent angular resolution and matched broadband X-ray polarimeters make it possible to design and build sensitive X-ray polarimetry missions. Recognizing the scientific opportunities afforded by X-ray polarimetry, NASA and ASI are currently developing the {\it Imaging X-ray Polarimetry Explorer (IXPE)} \citep{Weis:16}, a Small Explorer (SMEX) mission to be launched in 2021, which is expected to measure the polarization in the 2--8 keV energy range. 
In this white paper, we advocate for the science opportunities afforded by an {\it IXPE} follow-up mission, called the {\it X-ray Polarization Probe (XPP)} which promises a sensitivity improvement by a factor of 5--10 over {\it IXPE}, an energy bandpass broadened from 2--8 keV 
to 0.1--60 keV, and an angular resolution improvement from 30$''$ to 5$''$-10$''$ half power diameter (HPD). Such a mission will 
enable physics-type experiments with statistical samples of the most extreme objects in the Universe: black holes and neutron stars (see Table \ref{table}), and would allow us to probe matter, fields, and fundamental laws in extreme conditions. 

\begin{table}[b!]
 \caption{Summary of source classes (column 1),
 the fraction and number of sources
 per source class that can be observed with
 {\it IXPE} (column 2) and {\it XPP} (column 3) in 10$^5$\,s exposures with 
 a certain minimum detectable 
 polarization fraction ($<$10\% for magnetars and blazars, $<$5\% for all other source classes, 99\% confidence level) in a sample of bright RXTE sources, 
 as well as prominent science 
 drivers (column 4).\\[-7ex] \label{table}}
 \begin{center}
    \begin{tabular}{|p{2.5cm}|p{1.8cm}|p{1.8cm}|p{8.7cm}|} 
\hline 
\textbf{Class} & \textbf{\it IXPE} & \textbf{\it XPP} & \textbf{Science Drivers}\\
      \hline
      Stellar BHs &  
      73\% (27) & 89\% (33) & Accretion Disk Dynamics, 
      Role of \textbf{B}-field\\
	  AGNs &  0\% (0) & 52\% (15) & Black Hole Spin \& Inclination, 	   Corona Geometry\\
      Pulsars &  41\% (16) & 95\% (37)& QED Birefringence \\ 
      Magnetars & 0\%(0) & 57\% (8) & Stellar Surface, Plasma \& QED Birefringence \\
      HMXBs & 19\% (8) & 88\% (37) & Accretion, Plasma \& QED Birefringence \\   
      SNR+PWN &  4.5\% (1) & 77\% (17) & Origin of Cosmic Rays, Rel. Particle Acc. \\
      Blazars &  6\% (1) & 53\% (9) & 
      Jet Structure,
      Particle 
      Acceleration \\ \hline
    \end{tabular}
  \end{center}
\end{table}

Black holes and neutron stars emit a substantial, in most cases even dominant, fraction of their energy in the X-ray band, making this the preferred band for their study. A series of X-ray imaging, spectroscopy and timing missions led to spectacular insights about the nature of these (and other) cosmic X-ray sources, as recognized by Riccardo Giacconi's share of the 2002 Nobel Prize in Physics. {\bf X-ray polarimetry is expected to give another boost to the field by adding two more observables, the polarization fraction and angle, which encode vital information on the geometry of the systems. Importantly, the polarization constrains not only the geometry of {\it the source on femto-arcsecond scales} but also the {\it structure} of the magnetic and gravitational fields}. 
The measurement of the X-ray polarization will allow us to test our models of the X-ray emission mechanisms, 
the propagation of the X-rays through the curved spacetime in the vicinity of the compact objects,
the competition of plasma and vacuum birefringence, 
and the nature of X-ray scattering and selective, polarization dependent X-ray absorption 
\citep[e.g., see the reviews of][]{Mesz:88,Lei:97,Weis:09,Bell:10,Kraw:11}. \begin{figure*}[t!]
\begin{center}
\begin{minipage}{.4\linewidth}
\includegraphics[width=5.cm]{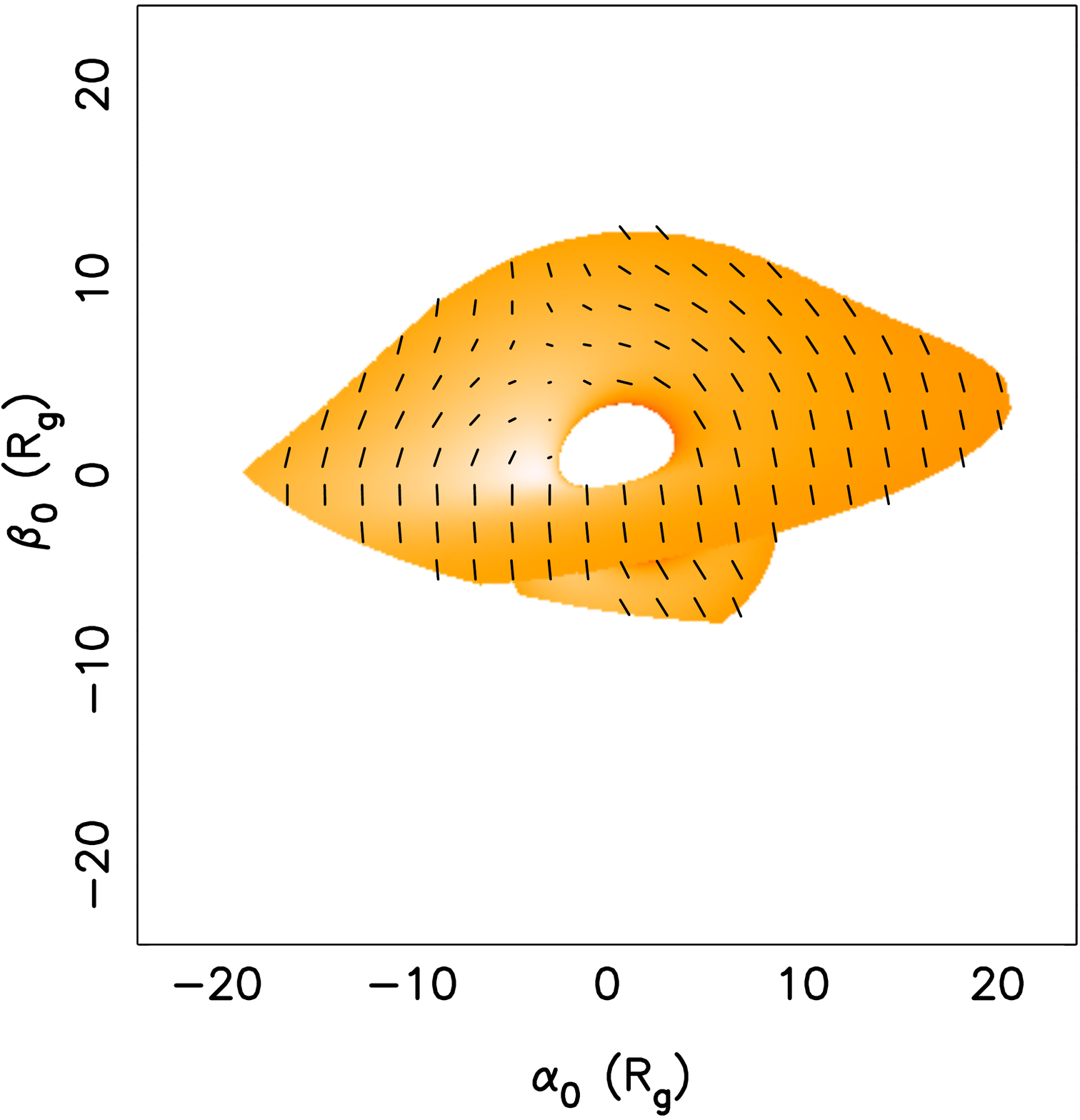}
\end{minipage}
\begin{minipage}{.3\linewidth}
\vspace*{-.4cm}
\includegraphics[width=5.cm]{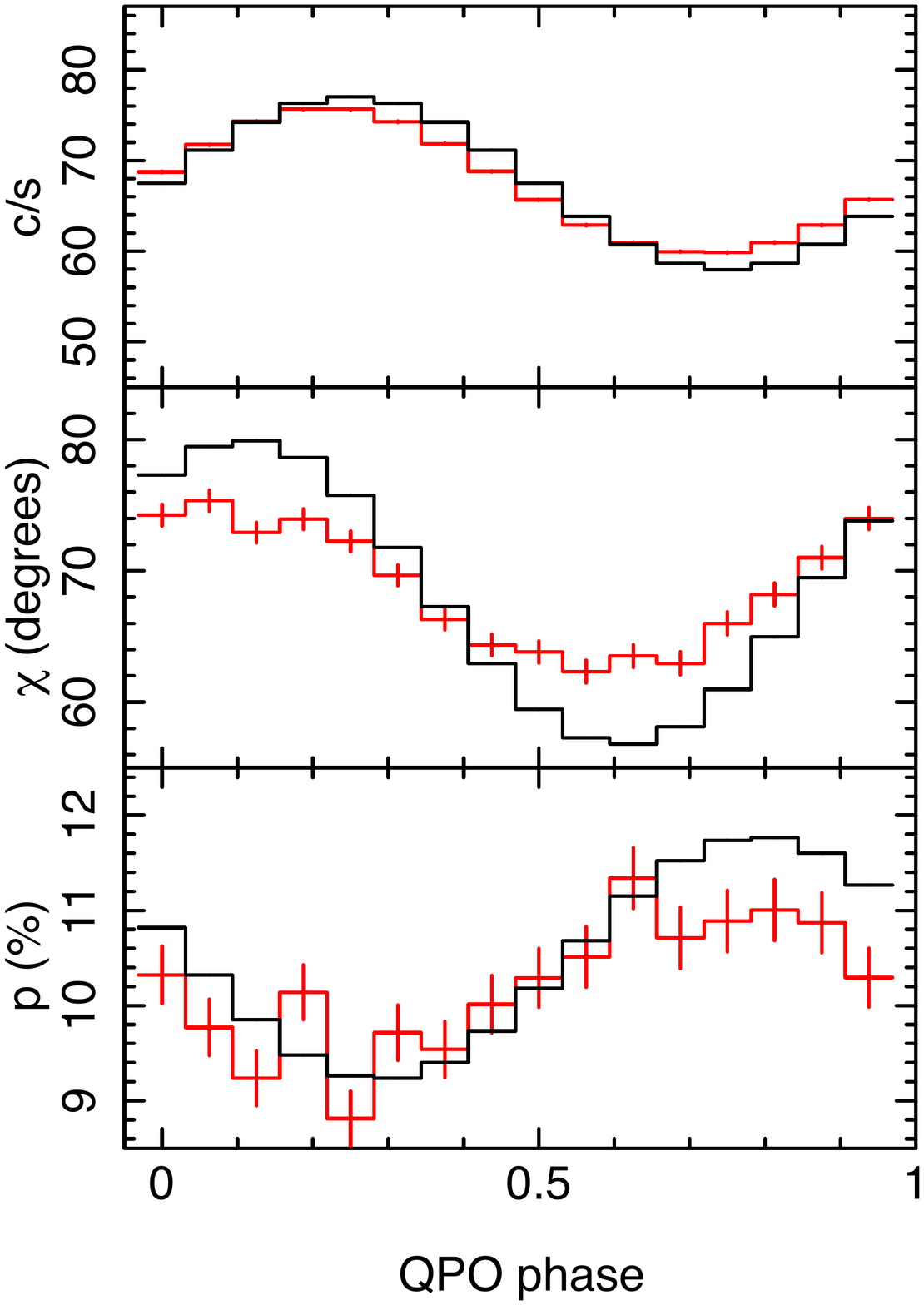}
\end{minipage}
\vspace*{-1.2cm}
\end{center}
\caption{\label{grs1915} High-throughput X-ray polarimetry will allow us to test and 
refine models of how black holes accrete matter and will open the new and exciting field 
of accretion disk dynamics. The left panel shows the polarization fraction and angle (encoded in black bars) of the X-rays of a precessing 
accretion disk. The right panel shows that XPP could reconstruct the precession based on a 10 hour
observation of the stellar mass black hole GRS 1915+105
\citep[from][]{Ingr:15}.}
\vspace*{-0.5cm}
\end{figure*}
{\it IXPE}, to be launched in 2021, will be the first dedicated X-ray polarimetry mission following the pioneering OSO-8 mission 
\cite[e.g.][]{Weis:78}, various satellite-borne mission with some
polarization sensitivity \cite[i.e.\ INTEGRAL, ASTROSAT, and Hitomi][]{Dean:08,Vada:15,Hito:18} 
and purpose-built balloon-borne hard X-ray missions \cite[i.e.\ PoGO+ and X-Calibur][]{Frii:18,Kisl:18}.
{\it IXPE} will deliver high signal-to-noise polarimetric measurements of the brightest Galactic sources and first results for a few extragalactic sources.
In the following, we will highlight several exceptionally promising science opportunities afforded by a {\it XPP}-type follow-up mission.\\[2ex]
{\bf Dissect the structure of inner accretion flow onto black holes and observe strong gravity effects:}
Black hole mass accretion is one of the most consequential processes in astrophysics, with mass accretion onto supermassive black holes generating a significant fraction of the entire electromagnetic energy output of the Universe, 
and playing an important role in the evolution of galaxies and galaxy clusters through feedback \citep[e.g.][]{Fabi:12,Silk:13,Chon:16}.  
X-ray polarimetry will provide qualitatively new ways to measure the intrinsic properties of black holes: their mass and angular momentum (spin) \citep{Li:09,Schn:09}. The distributions of these properties diagnose the formation and subsequent evolution of the population of BHs. 
For stellar-mass black holes in X-ray binary systems (XRBs), this informs on their formation via gravitational collapse of massive stars and on the evolution of the binary system. 
For supermassive black holes in active galactic nuclei (AGN), this informs on their growth throughout cosmological time, in tandem with their host galaxy, through accretion and mergers. The spins from X-ray observations (from Athena \citep{Nand:13} and/or a HEX-P-type mission \citep{Mads:18} using spin measurement methods developed based on the XPP data) will be an extremely valuable complement to supermassive black hole spins measured with the space-borne gravitational wave observatory LISA, and will be affected by completely different selection biases.
Although several independent methods have been developed to estimate stellar-mass black hole spins, their results disagree widely for some sources \citep[e.g.][]{Kraw:18}. 
X-ray polarimetry will offer an independent method to measure inclinations and black hole spins \citep[e.g.][]{Li:09,Dovciak:08}. 
Combining information from broadband X-ray polarimetry with higher-fidelity numerical simulations of accretion disks and the radiative processes can put spin measurement methods \citep{Zhan:97,Gier:01,McCl:14,Fabi:89,Bren:06,Mill:07,Bren:13} on a firmer footing.      

X-ray polarimetry will elevate accretion disk studies to a new level by observing the dynamics (precession, warping, accretion disk waves) 
of the accretion disks of mass accreting stellar mass black holes (Fig.\ \ref{grs1915}). 
For example, X-ray spectropolarimetric observations of low-frequency ($0.1-10$~Hz) Quasi Periodic Oscillations (QPOs) can give precision measurements of the Lense-Thirring precession of the accretion flow \citep{Ingr:15}.
Spectropolarimetric observations of high frequency ($>100$~Hz) QPOs would allow testing models such as the orbiting hot-spot model \citep{Schn:05,Behe:17}. 

Whereas Seyfert-type AGN are at the sensitivity threshold of {\it IXPE}, an {\it XPP}-like mission can increase the number of detected AGN 
from $\sim$5 to $>$20 with polarization energy spectra for bright sources enabling robust constraints on the geometry of the X-ray emitting corona.

Magnetic fields are thought to play a pivotal role in the accretion process: the magneto-rotational instability 
\citep[MRI][]{Balb:98} provides the effective viscosity required to transport angular momentum 
outwards and thus drive inward accretion of material, and vertically ordered magnetic fields are thought to launch jets \citep{Blan:77,Blan:82}.
Indeed, general relativistic magneto-hydrodynamic (GRMHD) simulations demonstrate that accretion and jet dynamics are crucially dependent 
on the strength and configuration of the magnetic field in the disk \citep{Nara:03}. 
A broadband polarization mission can constrain the strength and structure of the accretion disk magnetic field as the effects of Faraday rotation and QED (Quantum Electro-Dynamics) birefringence (see next section) may strongly depolarize the emission in a wavelength dependent way below 2~keV \citep{Davi:06}, and above 10~keV \citep{Caia:18}, respectively.\\[1.5ex]
{\bf Using neutron stars as fundamental physics laboratories:}
{Broadband X-ray polarimetry offers exciting new ways to investigate the physics and geometry of isolated neutron stars (INSs), magnetars and X-ray dim isolated neutron stars (XDINSs) {\it in particular}.} Furthermore, it enables the use of neutron stars as cosmic laboratories to quantitatively test QED high-field predictions, and to look for axion-like particles (ALPs).  
Magnetars, observationally identified with Soft Gamma-ray repeaters (SGRs) and Anomalous X-ray pulsars (AXPs), are characterized by ultra-strong fields (up to $20$--$30\,B_{\rm Q}$, {\it where $B_Q\sim 4.4\times 10^{13} \ \mathrm G$ is the quantum critical field}). Their (persistent) emission comprises a thermal component from the star surface ($kT\sim 0.5$--$1$~keV) plus a high-energy power-law tail reasonably produced by resonant Compton scattering onto magnetospheric currents 
\citep[RCS model, see][]{Thom:02,Bari:07,Nobi:08}.
\begin{figure*}[t]
\begin{center}
\begin{minipage}{7.5cm}
\includegraphics[width=7.3cm]{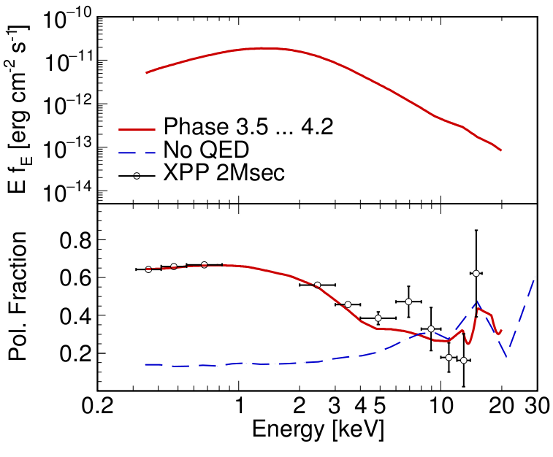}
\end{minipage}
\begin{minipage}{6cm}
\includegraphics[width=5.5cm]{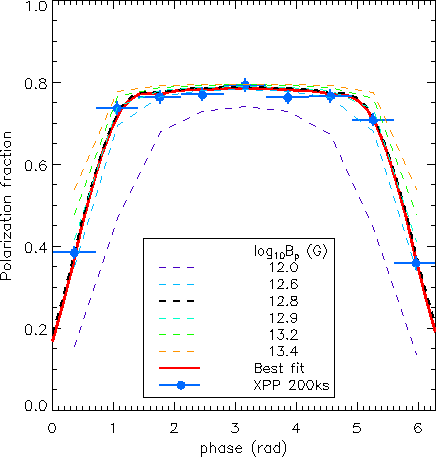}
\end{minipage}
\end{center}
\vspace*{-0.8cm}
\caption{Left: Simulated response of XPP for a 2 Ms observation of the AXP 1RXS~J170849.0--400910 (circles with error bars). 
The red curve shows the model from which mock data were produced \citep{Tave:14}. The figure shows one snapshot of a phase-dependent simulation for the phase interval 3.5--4.2 rad. Right: Simulated response of XPP for a 200 ks observation of the XDINS RX J1856.5--3754 (circles with error bars) obtained in the energy interval $0.42$--$0.59$ keV at different phases. The red curve shows the best-fit model; the dashed lines refer to models with different values of $B_{\rm pol}$.
\label{1708}}\vspace*{-1cm}
\end{figure*}
On the other hand, XDINSs are softer sources ($kT\sim 50$--$100$~eV) with thermal surface emission, and $B\approx B_{\rm Q}$. In such strong magnetic fields, photons {\it in vacuo} are linearly polarized in two normal modes, the ordinary (O) and extraordinary (X) ones. Below the electron cyclotron energy $\sim 11.6 (B/10^{12}\, {\rm G})$ keV, the X-mode opacity is suppressed with respect to the O-mode one so that the emerging thermal radiation is nearly 100\% polarized in the X-mode. 
While averaging the emission from regions with different local magnetic field directions tends to reduce the net polarization, 
the effect of QED vacuum birefringence \citep{Heis:36} can lock the polarization vector to the magnetic field direction 
up to large radii, where the field structure is more uniform \citep{Adle:71,Heyl:97}, leading to polarization fractions 
as high as 40\% to 80\% {\it at the detector} \citep{Heyl:03}.
{The left panel of figure \ref{1708} shows simulated data for the magnetar 1RXS~J170849.0--400910 
{\it according to} the RCS model \citep{Tave:14}. 
The observations below {\it IXPE's} 2~keV threshold can reveal the high polarization of the unscattered thermal photons. 
A 2 Ms $0.3$--$20$ keV observation can show a distinct decrease of the polarization fraction with the photon energy, 
a powerful probe of the RCS scenario.}

{In the case of INS sources the extended energy range of an {\it IXPE} follow-up mission can scrutinize the polarization in the soft (0.1--2 keV) X-ray band. 
Figure \ref{1708} (right) shows a simulated 200 ks XPP observation of the XDINS RX J1856.5-3754 assuming surface blackbody emission 90\% polarized in the X-mode \citep{Gonz:16}. 
The dipolar magnetic field $B_{\rm pol}$ and the angles {\it that the line-of-sight  and the magnetic axis make with the rotation axis ($\chi$ and $\xi$ respectively) } are those inferred from observations \citep[][and references therein]{Mign:14}. The full red line in the plot represents the best-fitting model obtained leaving $B_{\rm pol}$, $\chi$ and $\xi$ free to vary, while the dashed lines show models with different values of $B_{\rm pol}$. The input model parameters are recovered by the best fit with remarkable accuracy (typical error a few percent for the angles and $\sim 20\%$ for $B_{\rm pol}$). Therefore, a polarization measure below $2$ keV will measure the stellar magnetic field and geometry.}

Spectropolarimetric pulsar observations allow searches for ALPs motivated by the strong CP problem \citep{Pecc:77} 
and recent developments in string theory \citep{Arva:09}. 
ALPs can be produced via nucleon-axion bremsstrahlung in the core of a neutron star \citep{Grah:15}. 
The production rate is typically taken to be sub-dominant to neutrino emission, so as not to alter standard cooling rates \citep{Belo:16}.
Once produced, these ALPs can convert to photons in the magnetosphere through ALP-photon coupling via the Lagrangian term $g_{a\gamma} a \tilde{F}_{\mu\nu}F^{\mu\nu}$, and the conversion almost entirely occurs at a radial distance larger than the polarization radius.
The photons produced from ALP conversion are polarized parallel to the plane containing the electric field and the external magnetic field of the neutron star. The polarization can be computed by performing a scan over ALP masses, couplings, and energies, as well as typical values for the neutron star parameters \citep{Fort:18, Fort:18b}. For values allowed by the CAST experiment -- $g_{a\gamma} = 5 \times 10^{-11}$ GeV$^{-1}$, $m_{a} = 10^{-5}$ eV, $10$ km magnetar radius and $20 \times 10^{14}$ G surface magnetic field  -- the ALP-induced parallel polarization can lead to a clear ALP detection.

\begin{figure*}[t!]
\begin{minipage}{.47\linewidth}
\hspace*{1.2cm}
\vspace*{-0.2cm}
\includegraphics[width=7.6cm]{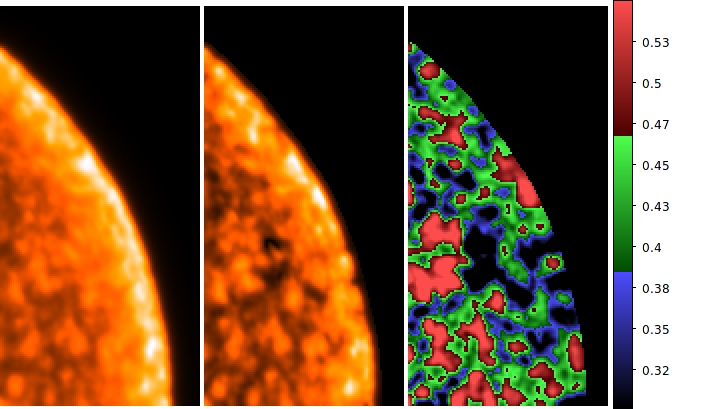}
\end{minipage}
\hspace*{1.0cm}
\begin{minipage}{.47\linewidth}
\vspace*{0.2cm}
\includegraphics[width=7.1cm]{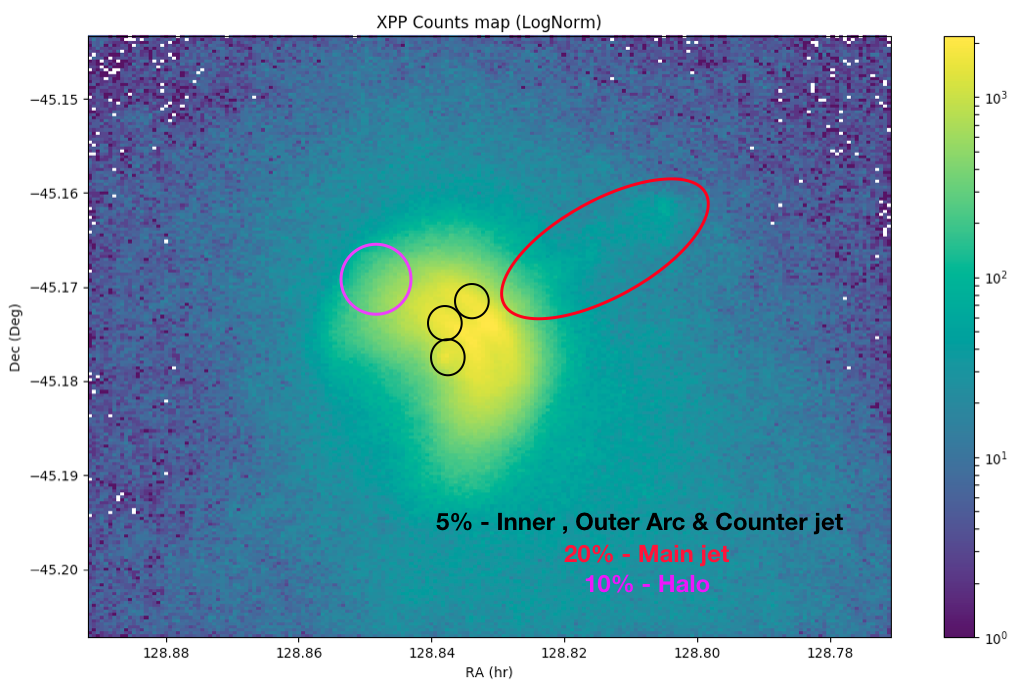}
\end{minipage}
\vspace*{-0.4cm}
\caption{\label{SNR} A next-generation X-ray polarimetric observatory can study the highly amplified magnetic turbulence thought to be key for the efficient acceleration of cosmic rays in supernova remnants and PWNs. 
The left panels show X-ray maps of a supernova shell simulated for a point spread function with 5$''$ HPD (left: intensity, center: polarized flux (intensity times polarization fraction), right: polarization fraction as  indicated by the color bar, see \citep{Byko:09}).
The right panel shows the accuracy (2 $\sigma$ confidence intervals) with which 
an {\it XPP}-type instrument can measure the 
linear polarization fraction from different regions of the Vela PWN in a 100 ksec observation.}
\vspace*{-0.4cm}
\end{figure*}
	A second-generation broadband X-ray polarimeter is ideally suited for observations of accretion-powered pulsars with known cyclotron resonant scattering features (CRSFs, absorption lines in the hard X-ray spectrum, seen between 10--80 keV), like Her X-1, Cep X-4, and Vela X-1. The CRSF energy is a direct tracer of the magnetic field strength, but the polarimetric observations  will allow us to probe deeper into the geometry of the accretion column and magnetic field. CRFS models currently under development \citep[e.g.][]{Schw:17} will allow us to compare polarization measurements with theoretical expectations from different geometries and to distinguish between a pencil and fan beam approximation of the accretion column geometry \citep{Mesz:92}. Even for accretion-powered pulsars without known CRSFs (and therefore no directly measured magnetic field strength), the polarimetric observations will probe the magnetic field through shifts in the polarization angle as function of energy. Ideal sources are bright X-ray pulsars like 4U 1909+07 and SMC X-4.\\[1.5ex]
{\bf How do cosmic particle accelerators work and what role do magnetic fields play?}
X-ray polarimetric observations are expected to lead to transformational discoveries about the mechanisms accelerating particles in supernova remnants (SNRs), pulsar wind nebulae (PWNe), pulsars, and in black hole jets.  
SNRs are believed to be the primary accelerator of galactic cosmic rays and the detection of the non-thermal radio to gamma-ray emission from SNRs is consistent with this paradigm.
X-ray polarimetry can inform us about turbulent magnetic field amplification in the particle accelerating shocks, a key mechanisms of the diffusive shock acceleration process. 
Indeed, high-resolution Chandra observations have recently revealed narrow synchrotron filaments in sources such as SN 1006 and Cas A. 
The TeV electrons emitting the synchrotron emission cool rapidly and thus trace the physical conditions in the surroundings of the shocks.   
As another fortunate coincidence, Faraday rotation, being proportional to the photon wavelength squared, partially depolarizes the radio emission, but not the X-ray emission. 
The polarimetric characterization of the magnetic turbulence requires angular resolutions of 10$''$ HPD  (see Fig. \ref{SNR}, left panel and \citep{Byko:09}). 

PWNe are strong synchrotron emitters characterized by broad-band continuum spectra\citep{Gaen:06}.  They are one of the most efficient relativistic accelerators in the Galaxy and the only accelerators for which we can image relativistic particle accelerating shocks. 
PWNe thus constitute a laboratory for the study of particle acceleration by relativistic shocks. There is good evidence that radio and X-ray particles are accelerated by different mechanisms, and that turbulence in the PWNe powers this process \citep{Tang:12,Zrak:17}.
Recent PIC simulations \citep{Comi:18}
indicate that non-thermal particles can be accelerated in turbulent and highly unstable MHD flows. 
Modeling of the surface brightness of the Crab wisps and torus and the Vela arc suggests that up to 50\% of the magnetic energy might be in a turbulent component \citep{Bucc:17}. 
Observations with a next-generation instrument can resolve the polarization of the emission from different regions (Fig.\ \ref{SNR}, right panel) and can test these predictions. 

The recent IceCube neutrino detection during the TXS~0506 flare strongly suggests that blazars can be the origin of ultra-high-energy cosmic rays \citep{IceC:18}. 
Whereas hadronic blazar models predict that the X-ray emission is highly polarized synchrotron emission ($10-20\%$ pol.\ fractions) from primary protons and/or from secondary $e^{+/-}$ pairs, leptonic models predict much lower pol. fractions \citep{Kraw:12,Kraw:13,Zhan:13,Zhan:16,Pali:18}.
A second-generation X-ray polarimeter will be able to distinguish the leptonic and hadronic radiation mechanisms by measuring the polarization of a large number of Low-Synchrotron-Peaked blazars which are too dim for {\it IXPE}. 
For the X-ray loud High-Synchrotron-Peaked blazars, a second-generation X-ray polarimeter can probe time-dependent polarization signatures, and shed light on the magnetic field evolution in the blazar zone, where the most extreme energy dissipation and particle acceleration take place and give rise to the X-ray emission \citep{Zhan:14,Tave:18}.
    %
\\[1.5ex]
{\bf Technology readiness:} An {\it XPP}-type mission uses technologies which are 
at a high readiness level, or will be by the time of a call for 
probe-sized missions. Multilayer coated X-ray mirrors can be fabricated based on  mono-crystalline Si  
\citep{Zhan:18} or replicated electroformed-nickel shells \citep{Gask:18}.
The medium-energy polarimeters and the high-energy polarimeter have been designed, built and tested in relevant environmental conditions for 
the {\it PRAXyS} \citep{Iwak:16}, {\it IXPE} \citep{Weis:16} and {\it XIPE} \citep{Soff:16}, 
and {\it NuSTAR} \citep{Harr:13} and {\it X-Calibur} \citep{Kisl:18} missions, respectively. 
The technology for the soft X-ray polarimeter is currently tested in a laboratory setting \citep{Mars:17}. 
Apart from further maturing these technologies, no new development are necessary, and the risks
of an {\it XPP}-type mission are exceptionally small.

\pagebreak

\end{document}